\documentclass[11pt,twoside]{article}

\usepackage{baaa-eng}
\usepackage{graphicx}
\usepackage{subfigure}
\usepackage{psfrag}
\usepackage{amssymb}

\usepackage[latin1]{inputenc}

\usepackage[T1]{fontenc} 
\usepackage{ae,aecompl} 

\usepackage{latexsym}
\usepackage{verbatim}
\usepackage[colorlinks=true,dvips]{hyperref}

\begin{document}
\vskip 1.0cm
\markboth{ R. Gamen et al}%
{Spectroscopic monitoring of Southern Galactic O and WN stars}

\pagestyle{myheadings}
\vspace*{0.5cm}
\parindent 0pt{ORAL PAPER}
\vskip 0.3cm

\title{Spectroscopic monitoring of Southern Galactic O and WN stars:
State of the Art in 2007}

\author{R. Gamen$^{1}$, 
R. Barb\'a$^{2}$,
N. Morrell$^{3}$,
J. Arias$^{2}$,
J. Ma\'{\i}z Apell\'aniz$^{4}$,
A. Sota$^{5}$,
N. Walborn$^{6}$,
E. Alfaro$^{4}$
}

\affil{%
(1) Complejo Astron\'omico El Leoncito - CASLEO\\
(2) Universidad de La Serena, Chile\\
(3) Las Campanas Observatory, The Carnegie Observatories, Chile\\
(4) Instituto de Astrof\'{\i}sica de Andaluc\'{\i}a, Espa\~na\\
(5) Universidad Aut\'onoma de Madrid, Espa\~na\\
(6) Space Telescope Science Institute, USA\\
}

\begin{abstract} 
We are conducting a spectroscopic monitoring of O and WN
-type stars in our Galaxy. 
In this work, we summarize some of our first results related to the
search for radial-velocity variations indicative of orbital motion. 
\end{abstract}

\begin{resumen}
Estamos llevando a cabo un monitoreo espectrosc\'opico de estrellas tipo O y WN de
nuestra Galaxia. 
En este trabajo, mostramos algunos de nuestros primeros resultados 
relacionados con la b\'usqueda de velocidades radiales que indiquen movimiento 
orbital.

\end{resumen}

\section{Introduction}

Stars with spectral type O and Wolf-Rayet (WR) are important objects
that play a crucial role in the dynamic and chemical evolution of galaxies.
They are the major source of ionizing radiation and, through their huge
mass-loss rates, they have a strong mechanical impact on their surroundings.
Despite their importance, our knowledge of these objects and of their
evolution is still fragmentary.
The parameters that predominantly determine the evolution of a massive
star are its mass, mass-loss rate, and perhaps its rotation. 
In this context, massive 
binaries with O or WR components are key objects because
their binary nature allows us to determine minimum masses from the
radial-velocity (RV) orbital solution and, if combined with techniques
to extract the orbital inclination, absolute masses.

Multiplicity introduces uncertainties in the determination of
massive star parameters and can dramatically affect their evolution
in close binaries.
It also causes biases in the determination of the
IMF of clusters (Ma\'{\i}z Apell\'aniz et al., 2005), and
provides clues to understand the origin of massive stars
(Zinnecker \& Yorke 2007).
Most studies about multiplicity of OB stars were
carried out through RVs and light curves analysis for short- and
medium-period systems; and speckle, high-angular resolution images,
dust-formation episodes, etc., for very long periods
(see also Zinnecker \& Yorke, 2007 and references therein).

\begin{figure}[!t]
  \centering
  \includegraphics[width=.99\textwidth]{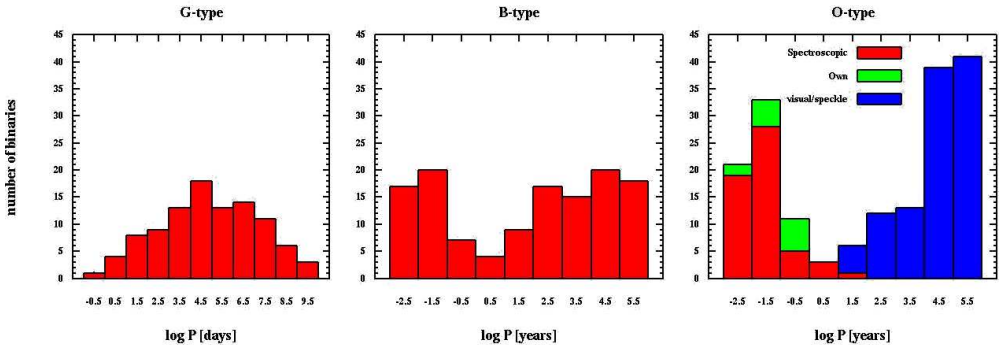}
  \caption{ \small
Period distribution among
G-type binaries (left), 
B-type binaries (center), and 
O-type binaries (right) 
showing the 13 new binaries discoveried by this project.
The dearth of O binaries between $\sim$ 30 days and 1000 years is remarkable
but note how our discoveries tend to fill the gap.
}
  \label{3histos}
\end{figure}

Quantitatively, in the Ninth Catalogue of spectroscopic binary
orbits (Pourbaix et al., 2004) there are 2420 binaries,
but only 78 with periods between 200 days and 1 year, 
with only one O-type star (O9.5V) among them.
It is also interesting to compare
the distribution of periods among different spectral types,
Mason et al. (1998) showed that among O stars it
presents a bimodal structure, Abt \& Cardona
(1984) showed that the even nearer B-type stars follow a much
flatter distribution (Figure~\ref{3histos}),
but Duquennoy \& Mayor (1991) found an unimodal Gaussian-like distribution
among solar-type stars. The bimodality for O stars is
likely an instrumental effect: spectroscopic surveys tend to detect
systems with short periods while, at the larger typical distance of
O stars, high-spatial-resolution ground-based surveys tend to detect
only systems with long periods. 

Most of the optical spectroscopic information about O-type stars available today
comes from old sources, and in many cases have been obtained through photographic
observations. The currently published atlases and digital spectrograms of O stars 
(e.g. Walborn \& Fitzpatrick 1990, 2000; Walborn et al. 2002), 
though detailed and descriptive, cover only a few tens of stars each. 

\section{The project}

Taking into account the above considerations, i.e. the importance of massive
stars as major modifiers of the ISM, the very dramatical role of the
multiplicity in their evolution, the apparent lack of long period
binary systems (P $\sim$ 1 year), and the dearth of available
complete high-quality spectroscopic atlas of O stars, we
are carrying out a long-term spectroscopic monitoring of
Galactic O and WN stars, with the main goals:

(a) To search for RV variations indicative of orbital motion in 
the stars for which there is no indication
of multiplicity in the Galactic O Stars catalog (GOS; Ma\'iz Apell\'aniz et al.,
2004) or in the VIIth Catalogue of Galactic Wolf-Rayet Stars (van der Hucht, 2000).
We aim to weight stellar masses applying Kepler's laws to binary
systems (when possible, using techniques to derive the orbital inclination);
(b) To establish a first epoch of uniform RVs for all the sample,
suitable for future searchings for spectroscopic binaries;
(c) To improve the spectral classification of most of the stars in the
sample and to compare them with those originally published 30-40 years ago to study
long-term variability related to evolutionary processes (e.g. LBV
behavior), changes due to the periastron passage of an unseen companion in a
very eccentric and large orbit,
high-energy phenomena (transient, X- or Gamma- ray burst), etc;
(d) To generate the
most-complete ever observed homogeneous library of O-type spectra
which will be available through the GOS catalog web page.

\section{The database}

The Southern part of this monitoring was started in 2005,
and so far we have collected about 1500 spectra of 144 O and WR stars,
using La Silla (ESO), and Las Campanas Observatories, in Chile, and 
Complejo Astron\'omico El Leoncito\footnote{CASLEO is operated under agreement
between CONICET, SECYT, and the National Universities of La Plata, C\'ordoba
and San Juan, Argentina.}, Argentina, facilities.
This spectroscopic dataset, is also
enlarged with spectra obtained during the last 10 years in CASLEO, and
some more retrieved from the ESO and AAO archives.
The Northern part of this project is being carried out from the Sierra 
Nevada and Calar Alto observatories, Spain, where we have obtained the 
spectra of 160 stars. 

\section{First results}

\begin{figure}[!t]
  \centering
  \includegraphics[width=.44\textwidth,angle=-90]{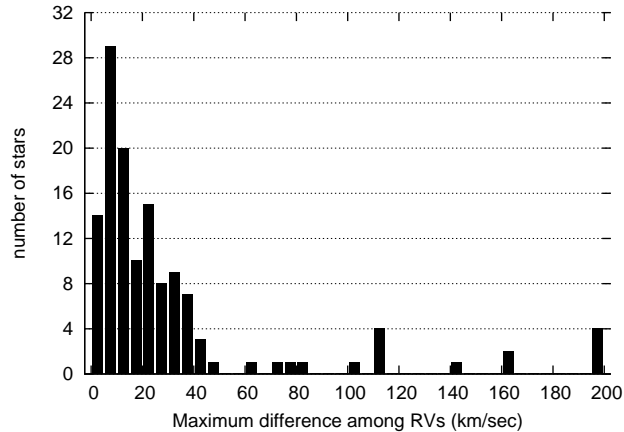}
  \caption{ \small
Histogram of the maximum difference among RVs
measured in the observed stars of our sample. Note that many stars
present RV variations larger than 10 km~s$^{-1}$, which is two-three times
larger than the maximum error we expect in RV measurements.
}
  \label{histoM-n}
\end{figure}

To date, we have observed 120 O- and 24 WN- type stars.
We discovered 13 binary systems with orbital periods
spanning from 2 to 250 days (See Fig.~\ref{3histos}),
2 stars present double-lined spectrum, but no reliable period 
was found,  
and nearly 80 stars show RV variations greater than
10 km~s$^{-1}$ for which periods have not yet been determined (see
Fig.~\ref{histoM-n}).

Among the new binaries we briefly can mention the
double-lined O+OB binary systems
HD~115455 
($P\sim 15.1 d$; 
$M_1 sin^3 i \sim  3 M_\odot$;
$M_2 sin^3 i \sim  1.5 M_\odot$)
and
HD~161853 
($P\sim 2.7 d$; 
$M_1 sin^3 i \sim  14 M_\odot$;
$M_2 sin^3 i \sim   6 M_\odot$).
But special attention deserves the massive binary systems
cl HM1 \#8, and HD~150135 (preliminary orbits published by Gamen et al. 2007),
WR~25, and WR~21a.

WR~25 was unveiled as a binary system (Gamen et al. 2006), and when 
was observed during the predicted quadratures, He {\sc ii} absorption lines 
from an hitherto unknown O-type companion were discovered, 
suggesting a ratio of the
semi-amplitudes ($K_{\rm O}/K_{\rm {WR}}$) 
at least larger than a factor 2 (See more in Gamen et al. 2007b).

WR~21a was also unveiled as a very massive WN+O binary.
We used the He~{\sc ii} $\lambda$4686 emission line and
the He~{\sc ii} $\lambda$5411 absorption line
to follow the motion of the WR and O component, respectively and
obtained $M_{\rm WR}$$\sim$87M$_\odot$ and $M_{\rm O}$$\sim$53M$_\odot$.
Although we prefer to be cautious, the WR component in the WR~21a
system could be the most massive star ever weighed in our galaxy,
``record'' held so far by both WN6ha components in the binary system WR~20a
(83+82$M_\odot$; Rauw et al, 2004; Bonanos et al, 2004)
located about 16\arcmin\, from WR~21a.

\small

\acknowledgements
We thank the directors and staff of CASLEO, LCO, and La Silla for the use of
their facilities. We also acknowledge the use at CASLEO of the CCD and data
acquisition system partly financed by U.S. NSF grant AST-90-15827 to
R. M. Rich. 
RG thanks the LOC for partially support the meeting costs.
RHB thanks support from FONDECYT Project No. 1050052.
JA acknowledges support from Fondo ALMA CONICYT No 31050004.

\end{document}